\documentclass[12pt]{article}
\usepackage{graphics}
\usepackage{epsfig}
\usepackage{amsmath}

\begin{document}

\setlength{\textheight}{240mm}
\voffset=-15mm
\baselineskip=20pt plus 2pt
\renewcommand{\arraystretch}{1.6}

\begin{center}

{\large \bf  The Einstein and M{\o}ller energy-momentum complexes in post-Newtonian approximation}\\
\vspace{5mm}
\vspace{5mm}
I-Ching Yang  \footnote{E-mail:icyang@nttu.edu.tw}

Department of Applied Science, National Taitung University, \\
Taitung 95002, Taiwan (R.O.C.)\\

\end{center}
\vspace{5mm}

\begin{center}
{\bf ABSTRACT}
\end{center}
In the first and second post-Newtonian approximation of the Schwarzschild metric, I obtain the energy 
component of the Einstein and M{\o}ller energy-momentum complex. Both energies involve the 
rest-mass energy $m$, the energy stored in the configuration and that in the gravitational field, but the 
energies of Schwarzschild spacetime in the Einstein and M{\o}ller prescriptions are the total 
mass-energy $M$. First, for general relativity, the rest-mass energy $m$ in the flat spacetime behaves 
like the bare mass, and the total mass-energy $M$ in the curved spacetime behaves like the experimentally 
observed mass. Second, the zero-potential surface is important condition for defining the energy of 
gravitational field, and plays an important role in the energy-momentum localization of 
general relativity.

\vspace{5mm}
\noindent

\newpage

In the evolution of physics, its unity is maintained by a network of correspondence principles, through 
which simpler theories maintain their vitality by links to more sophisticated but more accurate ones.  
The correspondence between the newer theory and its predecessor gives one the ability to recover the 
older theory from the newer. In view of Newton's theory of gravitation which is inconsistent with 
special relativity, Einstein was published general relativity based on equivalence principle, where the 
gravitational field is described by a metric. Furthermore, general relativity will reduce to Newton's 
theory of gravity in the limit of weak gravitational fields and low velocities. Einstein~\cite{E15a} 
built the Newtonian limit into general relativity, and also computed the precession of the perihelion 
which is one of the post-Newtonian effects. The formalism of Newtonian theory plus post-Newtonian 
corrections is called the ``post-Newtonian approximation"~\cite{MTW73}. 

One of the most important themes of general relativity is the energy-momentum localization which 
has no satisfactory solution. At present there is not an accepted definition of the localized 
energy-momentum associated with the gravitational field. Nevertheless, several approaches have been 
carried out to study the energy-momentum localization. A number of definitions for energy-momentum 
complexes in general relativity have been given by many authors, including Einstein~\cite{E15T}, 
M{\o}ller~\cite{M}, Tolman~\cite{T30},  Papapetrou~\cite{P48}, Bergmann and Thomson~\cite{BT53}, 
Landau and Lifshitz~\cite{LL62}, and Weinberg~\cite{W72}. Energy-momentum complexes are 
coordinate-dependent pseudotensorial quantities, except M{\o}ller, which can be used in quasi-Cartesian 
coordinates, more precisely in Schwarzschild Cartesian coordinates~\cite{SC}, in Kerr-Schild Cartesian 
coordinates~\cite{KSC}, and in generalized Painlev\'{e}-Gullstrand Cartesian coordinates~\cite{Y12}. 
Yang et al.~\cite{YR} have shown that the energy component of the energy-momentum complexes 
in Einstein prescription $E_{\rm E}$ and in M{\o}ller prescription $E_{\rm M}$ will be unlike beside 
vacuum case $T^0_0 = 0$ earlier. Then, Vagenas~\cite{V06}, Matyjasek~\cite{M08} and Yang~\cite{Y12} 
attempt to find out the relation formula about $E_{\rm E}$ and $E_{\rm M}$. In this article, according 
to the post-Newtonian approximation, I will study the difference between $E_{\rm E}$ and 
$E_{\rm M}$, and their relationship.

In order to obtain conserved quantities, Einstein introduce the energy-momentum complex
\begin{equation}
\Theta^{\mu}_{\nu} = \sqrt{-g} \left( T^{\mu}_{\nu} +t^{\mu}_{\nu}  \right),
\end{equation}
which satisfies the differential conservation form $\partial_{\mu} \Theta^{\mu}_{\nu} =0$. Here 
$T^{\mu}_{\nu}$ is the energy-momentum tensor of matter and $t^{\mu}_{\nu}$ is the 
energy-momentum pseudotensor from the gravitational field. Hence, an antisymmetric 
${\cal U}^{\mu \rho}_{\nu}$ in the indices $\mu$ and $\rho$ could be introduced mathematically as 
\begin{equation}
\Theta^{\mu}_{\nu} \equiv \frac{\partial {\cal U}^{\mu \rho}_{\nu}}{\partial x^{\rho}} ,
\end{equation} 
which is called ``superpotential". In the Einstein prescription, the energy-momentum complex is 
\begin{equation}
\Theta^{\mu}_{\nu} = \frac{1}{16\pi} \frac{\partial H^{\mu \rho}_{\nu}}{\partial x^{\rho}} ,
\end{equation}
with Freud's superpotential
\begin{equation}
H^{\mu \rho}_{\nu} = \frac{g_{\nu \sigma}}{\sqrt{-g}} \left[ (-g) \left( g^{\mu \sigma} g^{\rho \alpha}
- g^{\mu \alpha} g^{\rho \sigma} \right) \right] ,
\end{equation}
and in the M{\o}ller prescription, the energy-momentum complex reads 
\begin{equation}
\Theta^{\mu}_{\nu} = \frac{1}{8\pi} \frac{\partial \chi^{\mu \rho}_{\nu}}{\partial x^{\rho}} ,
\end{equation}
with M{\o}ller's superpotential
\begin{equation}
\chi^{\mu \rho}_{\nu} = \sqrt{-g} \left( \frac{\partial g_{\nu \alpha}}{\partial x^{\beta}} - 
\frac{\partial g_{\nu \beta}}{\partial x^{\alpha}}\right) g^{\mu \beta} g^{\rho \alpha} .
\end{equation}
Thus, the energy within the chosen region $\Sigma$ would be shown to be
\begin{equation}
E = \int_{\Sigma} \Theta^0_0 d^3x.
\end{equation}

In the weak-field case, for the solar system, the Schwarzschild metric can be expanded to give the first 
post-Newtonian form
\begin{equation}
ds^2 = (1 - 2\phi) dt^2 -(1 + 2\phi) (dx^2 + dy^2 + dz^2) .
\end{equation}
Here $\phi$ is the Newton's gravitational potential, and normalized such that $\phi (\infty ) =0$. 
To begin with, the corresponding Freud superpotential $H^{0i}_{0}$ is obtained 
\begin{equation}
H^{0i}_{0} = 4 \sqrt{\frac{1- 2\phi}{1+ 2\phi}} \partial_i \left( \frac{1+ 2\phi}{1- 2\phi} \right)  \frac{x^i}{r} ,
\end{equation}
and be expanded in powers of $\phi$ as
\begin{equation}
H^{0i}_{0} = \left[ 4 \partial_i \phi + {\cal{O}} (\phi^2) \right]  \frac{x^i}{r} .
\end{equation}
The energy component of the Einstein energy-momentum complex is evaluated to be
\begin{equation}
{}_{E} \Theta^0_0 = \frac{1}{4\pi} \nabla^2 \phi ,
\end{equation}
and these higher-order terms $ {\cal{O}} (\phi^2)$ is ignored. Eventually, in the Einstein prescription, 
the energy within the region $\Sigma$ that includes mass distribution is given by 
\begin{equation}
E_{\rm E} = \frac{1}{4\pi} \int_{\Sigma} \nabla^2 \phi d^3 x .
\end{equation}
Since the gravitational potential satisfes the field equation with a matter density $\rho$ 
\begin{equation}
\nabla^2 \phi = 4\pi \rho 
\end{equation}
and the total mass is defined by
\begin{equation}
\int_{\Sigma} \rho d^3 x= m .
\end{equation}
Hence, the energy in the Einstein prescription can be rewritten to
\begin{equation}
E_{\rm E} = m .
\end{equation}
Subsequently, the corresponding M{\o}ller superpotential is found to be
\begin{equation}
\chi^{0i}_{0} = - \sqrt{\frac{1+ 2\phi}{1- 2\phi}} \partial_i (1- 2\phi)  \frac{x^i}{r} ,
\end{equation}
which is expanded in power of $\phi$ as
\begin{equation}
\chi^{01}_{0} =  \left[ 2 \partial_i \phi + {\cal{O}} (\phi^2) \right]  \frac{x^i}{r} .
\end{equation}
The energy component of the M{\o}ller energy-momentum complex is described to
\begin{equation}
{}_{M} \Theta^0_0 = \frac{1}{4\pi} \nabla^2 \phi + {\cal{O}} (\phi^2) .
\end{equation}
After the higher-order terms ${\cal{O}} (\phi^2)$ is ignored, the energy in the M{\o}ller prescription 
with the region $\Sigma$ is exhibited
\begin{equation}
E_{\rm M} = m .
\end{equation}
The term of the right-hand side in Eq.(15) and Eq.(19) is the ``rest-mass energy" as the invariant mass 
is the rest energy in special relativity.

Next, the second post-Newtonian form of the Schwarzschild metric is considered for post-Newtonian
corrections to the Newtonian treatment~\cite{W72}
\begin{equation}
\begin{split}
ds^2 = & ( 1 -2\phi + 2\phi^2) dt^2 - (1 + 2\phi +\phi^2 ) (dx^2+dy2+dz^2)  \\
& - \frac{ \phi^2}{r^2} (x dx + y dy + z dz)^2 .
\end{split}
\end{equation}
Therefore, the Freud superpotential is given
\begin{equation}
H^{0i}_0 =  \left[ 4\partial_i \phi + 2\phi \partial_i \phi + \frac{2\phi^2}{r} + {\cal{O}} (\phi^3) \right]  \frac{x^i}{r} .
\end{equation}
The energy component of the Einstein energy-momentum complex is 
\begin{equation}
{}_{E} \Theta^0_0 = \frac{1}{4\pi} \nabla^2 \phi +\frac{1}{8\pi} [ \phi \nabla^2 \phi + (\nabla \phi)^2] 
+ \frac{1}{8\pi} \nabla \cdot \frac{\phi^2}{r} \hat{r} ,
\end{equation}
and the energy in the Einstien prescription can be shown 
\begin{equation}
E_{\rm E} = m + \frac{1}{2} \int_{\Sigma} \rho \phi d^3x + \frac{1}{8\pi} \int_{\Sigma} (\nabla \phi)^2 d^3x 
+ \frac{1}{8\pi} \oint_{\partial \Sigma} \frac{\phi^2}{r} \hat{r} \cdot d\vec{a}  .
\end{equation}
Here, the second term of Eq.(23) represents the energy stored in the configuration
\begin{equation}
E_{\rm config} \equiv \frac{1}{2} \int_{\Sigma} \rho \phi d^3 
\end{equation}
and the third term represents the energy stored in the gravitational field
\begin{equation}
E_{\rm field} \equiv \frac{1}{8\pi} \int_{\Sigma} (\nabla \phi)^2 d^3 x . 
\end{equation}
Thus, the energy in the Einstien prescription can be rewritten to
\begin{equation}
E_{\rm E} = m + E_{\rm config} + E_{\rm field} 
+ \frac{1}{8\pi} \oint_{\partial \Sigma} \frac{\phi^2}{r} \hat{r} \cdot d\vec{a} .
\end{equation}
Afterward the corresponding M{\o}ller superpotential is found 
\begin{equation}
\chi^{01}_{0} = [ 2 \partial_i \phi - 4\phi \partial_i \phi + {\cal{O}} (\phi^3) ] \frac{x^i}{r} .
\end{equation}
The energy component of the M{\o}ller energy-momentum complex is described as
\begin{equation}
{}_{M} \Theta^0_0 = \frac{1}{4\pi} \nabla^2 \phi - \frac{1}{2\pi} \left[  \phi \nabla^2 \phi 
+ ( \nabla \phi)^2  \right] + {\cal{O}} (\phi^3) 
\end{equation}
To ignore the higher-order terms ${\cal{O}} (\phi^3)$, the energy in the M{\o}ller prescription 
with the region $\Sigma$ is exhibited as 
\begin{equation}
E_{\rm M} = m - 4 E_{\rm config} - 4 E_{\rm field} 
\end{equation}

In summary, I obtain the energy of the Schwarzschild spacetime in the first post-Newtonain 
approximation by using the Einstein and M{\o}ller energy-momentum complexes, and the results 
can be rewritten as
\begin{equation}
E_{\rm E} = m + \left( E_{\rm config} + E_{\rm field} \right) 
+ \frac{1}{8\pi} \oint_{\partial \Sigma} \frac{\phi^2}{r} \hat{r} \cdot d\vec{a}
\end{equation}
and
\begin{equation}
E_{\rm M} = m - 4  \left( E_{\rm config} + E_{\rm field} \right) .
\end{equation}
First, both energies $E_{\rm E}$ and $E_{\rm M}$ involve the rest-mass energy $m$, the energy 
stored in the configuration $E_{\rm config}$ and that in the gravitational field $E_{\rm field}$ and 
all terms can be realized in special relativity and Newton's theory of gravity. However, in Schwarzschild 
spacetime, the energies in the Einstein and M{\o}ller prescriptions are
\begin{equation}
\bar{E}_{\rm E} = \bar{E}_{\rm M} = M ,
\end{equation}
which $M$ is the ``total mass-energy"~\cite{MTW73} . Referring to quantum electrodynamics, 
due to vacuum polarization, bare mass and bare charge of electrons are not measurable. Hence,
for general relativity, the rest-mass energy $m$ in the flat spacetime behaves like bare mass, and the 
total mass-energy $M$ in the curved spacetime behaves like the experimentally observable mass. The 
total mass-energy $M$ can be regarded as a modified $m$. Second, the difference of energy 
between the Einstein and M{\o}ller prescriptions~\cite{YR} is defined as
\begin{equation}
\Delta E = E_{\rm E} - E_{\rm M} ,
\end{equation}
giving 
\begin{equation}
\Delta E = \frac{5}{8\pi} \oint_{\partial \Sigma} \phi \nabla \phi \cdot d \vec{a} 
+  \frac{1}{8\pi} \oint_{\partial \Sigma} \frac{\phi^2}{r} \hat{r} \cdot d\vec{a} .
\end{equation}
While gravitational potential $\phi =0$ or gravitational field $\nabla \phi =0$ in the region 
$\partial \Sigma$, the difference of energy $ \Delta E$ will become zero. Let us turn our attention 
to the calculation of potential energy, the definition of zero-potential surface is primarily important 
condition. In the Einstein prescription, the first derivative of the metric $g_{\mu \nu , \alpha}$
was required to be zero, so the zero-potential surface of the Einstein energy-momentum complex is 
a flat spacetime. However, there is no manifest condition in the M{\o}ller prescription, and the 
zero-potential surface of the M{\o}ller energy-momentum complex can not be described. The 
zero-potential surface will be an important role in the energy-momentum localization of general 
relativity and should be considered seriously. Meanwhile, the choice of reference in the boundary 
term of the quasi-local energy expressions~\cite{CJN} and the role of thermodynamic 
potential~\cite{Y12, CNC}, could be associated with the quasi-localized energy.

\begin{center}
{\bf Acknowledgments}
\end{center}
I would like to thank Prof. Yaw-Hwang Chen for discussions and Mr. Yi-Hsun Li for partial 
calculation.

\end{document}